\documentclass[12pt]{spieman}  
\usepackage{amsmath,amsfonts,amssymb}
\usepackage{graphicx}
\usepackage{setspace}
\usepackage{tocloft}
\usepackage{lineno}

\usepackage{multirow}%
\usepackage{mathrsfs}%
\usepackage[title]{appendix}%
\usepackage{xcolor}%
\usepackage{textcomp}%
\usepackage{manyfoot}%
\usepackage{booktabs}%
\usepackage{algorithm}%
\usepackage{algorithmicx}%
\usepackage{algpseudocode}%
\usepackage{listings}%
\usepackage{changepage}

\newcommand{\hhl}{}

\title {MMIS-Net for Retinal Fluid Segmentation and Detection}
\author[a]{Nchongmaje Ndipenoch}
\author[a]{Alina Miron}
\author[a]{Kezhi Wang}
\author[a,*]{Yongmin Li}
\affil[a]{Department of Computer Science, Brunel University London, Uxbridge, UB8 3PH, UK}

\cftpagenumbersoff{figure}
\cftpagenumbersoff{table} 
\begin{document} 
\maketitle

\begin{abstract}
\begin{adjustwidth}{2em}{0pt} 
 \textbf{Purpose:} 
 Deep learning methods have shown promising results in the segmentation, and detection of diseases in medical images. However, most methods are  trained and tested on data from a single source, modality, organ, or disease type, overlooking the combined potential of other available annotated data. Numerous small annotated medical image datasets from various modalities, organs, and diseases are publicly available. In this work, we aim to leverage the synergistic potential of these datasets to improve performance on unseen data.\\
 \textbf{Approach:} 
 To this end, we propose a novel algorithm called MMIS-Net (MultiModal Medical Image Segmentation Network), which features Similarity Fusion blocks that utilize supervision and pixel-wise similarity knowledge selection for feature map fusion. Additionally, to address inconsistent class definitions and label contradictions, we created a one-hot label space to handle classes absent in one dataset but annotated in another. MMIS-Net was trained on 10 datasets encompassing 19 organs across 2 modalities to build a single model. \\
 \textbf{Results:} 
 The algorithm was evaluated on the RETOUCH grand challenge hidden test set, outperforming large foundation models for medical image segmentation and other state-of-the-art algorithms. We achieved the best mean Dice score of 0.83 and an absolute volume difference of 0.035 for the fluids segmentation task, as well as a perfect Area Under the Curve of 1 for the fluid detection task.\\
\textbf{Conclusion:} 
The  quantitative results highlight the effectiveness of our proposed model due to the incorporation of Similarity Fusion blocks into the network’s backbone for supervision and similarity knowledge selection, and the use of a one-hot label space to address label class inconsistencies and contradictions.\\
\end{adjustwidth}
\end{abstract}


{\noindent \footnotesize\textbf{*}Yongmin Li,  \linkable{yongmin.li@brunel.ac.uk} }

\begin{spacing}{1}   

\section{Introduction}
\label{sec:introduction}
Image segmentation is a widely studied problem in the deep learning community and is paramount in medical image analysis, diagnostics, and monitoring the progression of pathogens/diseases. Medical image segmentation tasks involve diverse modalities such as Optical Coherence Tomography (OCT), Computed Tomography (CT), Positron Emission Tomography (PET), Magnetic Resonance Imaging (MRI), Ultrasound, X-ray, and many more, incorporating various anatomical structures such as the retina, brain, neck, fetal tissues, chest, abdomen, cells, and more. Several small datasets with their corresponding annotations/labels from different modalities and anatomic regions are available in the public domain. This availability has sparked the development of numerous deep learning algorithms for lesion segmentation in medical imaging. However, most of these algorithms are typically trained on a single modality for a specific anatomic structure or problem, leading to challenges in generalization to new, unseen datasets like in real-world scenarios. One of the main causes of this issue is the high variability in image quality stemming from different modalities, collected across various medical centers using machines from different manufacturers and annotated by radiologists with varying levels of experience. One approach to circumventing this problem is to increase the diversity of the training set by combining images from various modalities, representing different anatomic structures, and collected across different medical centers using devices from various vendors. 
\hhl{Other approaches in the past that have combined data from multiple diverse sources include: a single network with a shared encoder and separate decoders for each dataset is presented in \cite{chen2019med3d}. Similarly, a single network across different domains using a common shared point-wise convolution and domain-specific adapters, where each domain adapter contributed to and shared knowledge from the shared point-wise convolution, is introduced in \cite{huang20193d}. Also,  a conditional network to segment multiple classes from a single dataset is proposed in \cite{dmitriev2019learning}. One limitation of these approaches is that they are designed for multi-organ segmentation and do not take into consideration overlapping targets (structures or organs that are labelled in one dataset but absent in another). }
To this end, we propose a novel algorithm: MMIS-Net (MultiModal Medical Image Segmentation Network), which combines Convolutional Neural Network and the Similarity fusion blocks to simultaneously segment lesions from different anatomic structures across diverse image modalities. Our main contributions are as follows:
1) We introduce MMIS-Net, a novel algorithm designed to train a single model to segment multiple lesions from various body structures across diverse image modalities simultaneously. MMIS-Net incorporates similarity fusion blocks into its architecture, utilizing supervision and pixel-wise selection knowledge for feature map fusion. This approach reduces irrelevant and noisy signals in the output.
2) We efficiently created a one-hot label space to address the inconsistent class definitions and label contradiction problem, covering diverse modalities and body regions in a multiclass segmentation problem.
This strategy effectively manages classes that are absent in one dataset but annotated in another during training. Also, it retains different annotation protocol characteristics for the same target structure and allows for overlapping target structures with different levels of detail, such as liver, liver vessels, and liver tumors.

The rest of the paper is organized as follows. A brief review of the previous studies is provided in Section \ref{section:background}. Section \ref{section:method} presents our method. The datasets, experiment with results and visualisation are presented in Section \ref{section:experiments}. Finally, the conclusion with our contributions, limitation and future work are described in Sections \ref{section:conclusions}. 

\section{Background}
\label{section:background}
In recent years, various deep learning approaches have been proposed for medical image segmentation, ranging from specific design models to large foundation models. Some of these will be briefly reviewed as follows, while more comprehensive reviews of recent work can be found in 
\cite{liu2021review,abdou2022literature,amiri2024deep,wang2022medical,fu2021review,qureshi2023medical,amiri2023resilient,amiri2023personal,amiri2024applications,amiri2024adventures, salazar2011optic,kaba2014retinal,salazar2010retinal,kaba2013segmentation,kaba2015retina,eltayef2016detection,salazar2012mrf,dodo2018graph,dodo2019retinal,huang2024segmenting,luo2025segrap2023,ehab2024unet}.

\subsection{Specific Design Algorithms }
\label{subsectionsection:sda}

The U-Net, a convolutional neural network (CNN) for biomedical image segmentation featuring encoder and decoder paths, along with a bottleneck, is introduced in \cite{ronneberger2015u}. The encoder path is utilized for capturing contextual information, while the decoder path is employed for localization. The Deep\_ResUNet++, an extension of ResUNet++, is introduced in \cite{ndipenoch2022simultaneous} for the simultaneous segmentation of layers and fluids in retinal OCT images. The algorithm was evaluated on the Annotated Retinal OCT Images (AROI) database \cite{melinvsvcak2021aroi}, achieving a Dice score of 0.9 and above for all eight classes on the test dataset. Another CNN architecture for retinal image segmentation is presented in \cite{ndipenoch2023retinal} for the segmentation of seven retinal layers and one fluid class. The algorithm was evaluated on the Duke Dataset \cite{chiu2015kernel}, achieving a mean Dice score of 0.77. The ReLayNet is introduced in \cite{roy2017relaynet} for the segmentation of retinal layers and fluid. The architecture employs CNN as a backbone in combination with a loss function comprising of weighted logistic regression and Dice overlap loss. The method was evaluated on the Duke dataset, achieving a mean Dice score of 0.75. The nnU-Net, a self-configuring method for deep learning-based biomedical image segmentation, which employs U-Net as a backbone is presented in \cite{isensee2021nnu}. The method was evaluated on 11 international biomedical image segmentation challenges, consisting of 23 different datasets and 53 segmentation tasks, and achieved first place in 33 out of the 53 tasks. Since the introduction of the nnU-Net, several of its variants have been proposed, including: \cite{ndipenoch2024performance, ndipenoch2023nnunet, mcconnell2023advanced, mcconnell2023exploring}.
The RETOUCH grand challenge was launched in 2017 for the segmentation of three retinal fluids from OCT images acquired from three device vendors: Topcon, Spectralis, and Cirrus. 
Top teams and algorithms published on the challenge website include \cite{retouchwebsite}:
IAUNet\_SPP\_CL \cite{xing2022multi}: This approach presented a combination of a graph-theoretic method, a fully convolutional neural network (FCN), curvature regularization loss function, and spatial pyramid pooling (SPP) modules using U-Net as the backbone.
SFU \cite{lu2017retinal}: A combination of a 3-part CNN-based framework and a Random Forest (RF) is introduced. The CNN is used for pre-processing and feature extraction, while the RF is used for pixel classification.
UMN \cite{rashno2017detection}: This method combines a CNN and a graph-shortest path (GSP) method. The authors used CNN to extract the region of interest (ROI), thereby reducing the training time, and GSP was used for pixel classification.
MABIC \cite{kang2017deep}: This approach introduced a double U-Net architecture concatenated in series. The first part is used to extract the ROI, which serves as input to the second part that is used for segmentation.
RMIT \cite{tennakoon2018retinal}: A combination of a deep neural network and an adversarial loss function is presented.
RetinAI \cite{apostolopoulos2017simultaneous}: A standard 2D U-Net with residual connections is presented.
An unsupervised technique for noise transfer in the domain adaptation of retinal OCT images using a noise adaptation approach based on singular value decomposition (SVDNA) \cite{koch2022noise} is introduced.

\subsection{Universal Algorithms }
\label{subsectionsection:universal}
The 3D $U^2$-Net, a 3D universal U-Net for
multi-domain medical image segmentation is introduced in \cite{huang20193d}. The CLIP-Driven, a universal model for organ segmentation and tumor detection is presented in \cite{liu2023clip}. The authors combined text and image datasets to simultaneously segment organs and detect tumors from fourteen datasets.
A multi-source domain generalization model based on domain and content adaptive convolution (DCAC) is proposed in \cite{hu2021domain}.
The MDViT, a multi-domain vision transformer for small medical image segmentation datasets, is proposed in \cite{du2023mdvit}. 
The MultiTalent, a multi-dataset approach for medical image segmentation, is presented in \cite{ulrich2023multitalent} for the segmentation of multiple CT datasets with diverse and conflicting class definitions. 

\subsection{Foundation Models  }
\label{subsectionsection:FM}
The Segment Anything Model (SAM), a foundation model for general image segmentation, developed by researchers at Meta, is introduced in \cite{kirillov2023segment}. The model is trained on 1 billion masks and 11 million images. Ever since the introduction of SAM, several of its variants tailored for medical image segmentation have been introduced, some of which will be reviewed as follows. SAM was initially trained on 2D images. The MA-SAM \cite{chen2023ma} fine-tuned SAM by incorporating 3D adapters into the transformer blocks of the image encoder, adding a crucial third dimension for 3D medical image segmentation tasks. 
SAMed, introduced in \cite{zhang2023customized}, applies the low-rank-based (LoRA) fine-tuning strategy \cite{hu2021lora} to the SAM image encoder. It fine-tunes SAMed together with the prompt encoder and the mask decoder on labeled medical image segmentation datasets.
Another approach that adopts the 2D SAM for 3D medical image segmentation is SAM-Med2D \cite{cheng2023sam}. \hhl{The 
SAMedOCT is presented in \cite{fazekas2023samedoct}. The authors adapted SAM for the segmentation of three retinal fluids on the RETOUCH challenge datasets \cite{bogunovic2019retouch}. SAMedOCT achieved the best AVD score of 0.033 for the PED class but was outperformed by MMIS-Net in all other classes for both the DS and AVD scores.}

\section{Method}
\label{section:method}

\begin{figure*}[h]
\centerline{\includegraphics[width=0.8\textwidth]
{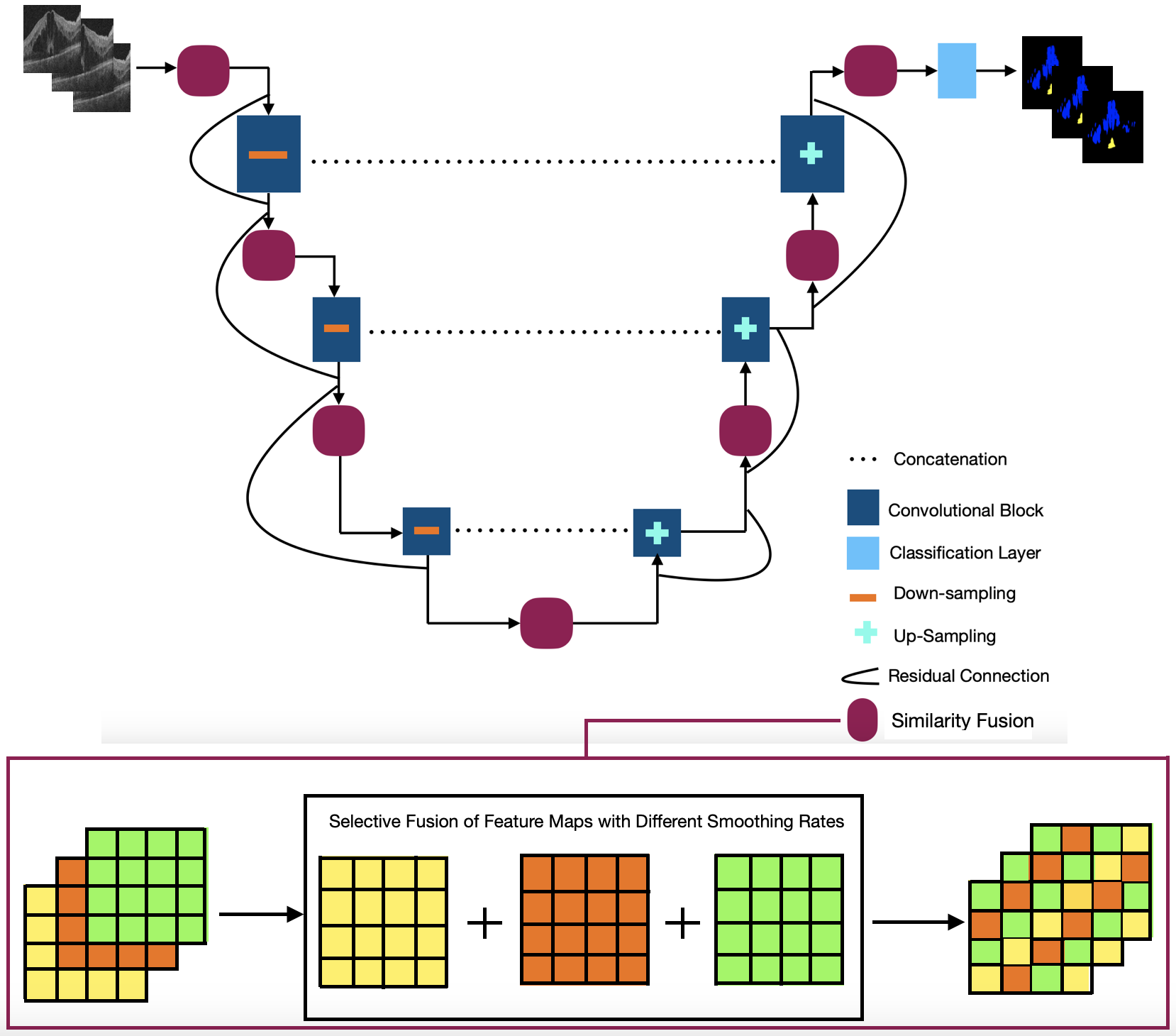}}
\caption{A high-level illustration of the MMIS-Net architecture demonstrating the contracting and expanding paths, residual connections, and the similarity fusion blocks. Further details of the fusion block, illustrating the feature map fusion using supervision and pixel-wise similarity selection of images at different smoothing scales, is shown at the bottom.}
\label{fig:MMIS-Net}
\end{figure*}

Given a dataset collection of \(K\) datasets \(D^{(k)}\), \(k \in [1, K]\), with \(N^{(k)}\) image 
\((x)\) and label pairs \((y)\)
\(D^{(k)} = \{(x, y)^{(k)}_1, \ldots, (x, y)^{(k)}_{N^{(k)}} \}\). Every pixel \(x^{(k)}_i\), \(i \in [1, I]\), is assigned to one class \(c \in C^{(k)}\), where \(C^{(k)} \subseteq C\) is the label set associated with dataset \(D^{(k)}\). 
We combined all the label images into a single one-hot label space for all the datasets and each class is assigned a unique label value as demonstrated in Tale ~\ref{tab:labels}. 
Combining partially annotated datasets presents its own challenges, and here are some:
1) Label Index Inconsistency: The same organ can be labeled with different indexes in different datasets.
2) Background Inconsistency: An organ is marked as background in one dataset but as foreground in another. For example, in the Pancreas-CT dataset \cite{nihPancreasCT}, the pancreas is marked as foreground, but it is marked as background in the MSD Spleen dataset \cite{medicaldecathlon}.
3) Absent of Organ Labels: The same organ is labeled in one dataset but absent in another dataset that also contains the organ. For example, in the MSD Liver dataset, both the liver and liver tumor are segmented. In contrast, in the MSD Hepatic Vessels dataset, the labeled targets are the vessels and tumors within the liver, but not the liver itself.
\hhl{4)Organ overlapping. There is overlap between
various organs. For example, Hepatic Vessel is part of
the Liver and Kidney Tumor is a sub-volume of the
Kidney.
Various methods, such as \cite{liu2023clip}, have tried to address these challenges by combining labels with text embedding and adopting a masked back-propagation mechanism. In this work, we use labels only and adapt the network architecture to effectively manage classes that are absent in one dataset but annotated in another during training. Incorporating text embeddings required two training branches, a text branch and a vision branch. The text branch adds an extra layer of complexity to the model.
Our strategy  does not use text embeddings and also retains different annotation protocol characteristics for the same target structure and accommodates overlapping target structures with varying levels of detail, such as the liver, liver vessels, and liver tumors.}
Even if classes from different datasets refer to the same target structure, we treat them as unique due to the unknown and potentially variable annotation protocols and labeling characteristics across datasets. Consequently, the network must be able to predict multiple classes for a single voxel/pixel to accommodate these inconsistent class definitions. To address the label contradiction problem, at the classification layer we decouple the segmentation outputs for each class by using a Sigmoid activation function instead of the commonly used Softmax activation function. The network shares the same backbone parameters \(\Theta\) but has independent segmentation head parameters \(\Theta_c\) for each class. The Sigmoid probabilities for each class are defined as \( \hat{y}_c = f(x, \Theta, \Theta_c) \). This modification allows the network to assign multiple classes to a single pixel, thus enabling overlapping classes and preserving all label properties from each dataset. Consequently, the segmentation of each class can be treated as a binary segmentation task.

The MMIS-Net (MultiModal Medical Image Segmentation Network) is composed of five main components: a contracting path (the encoder), an expansion path (the decoder), the similarity fusion block, residual connections, and a class-adaptive loss function.

\subsection{The Contracting Path}
\label{subsectionsection:contractingpath}
The contracting path is used to capture contextual information and as we go down the contracting path the image is halved after every convolutional block. Each block consists of two 3x3 convolutions followed by a ReLU (Rectified Linear Unit) activation function and next is followed by a 2x2 max-pooling, which reduces the feature map by half.

\subsection{The Expanding Path}
\label{subsectionsection:expandingpath}
The expanding path is used for pixel localization. As we go up the expanding path, the feature map is doubled after every convolutional block by concatenating the feature map of the expanding path with its corresponding map in the contracting path. Each block in the expanding path is composed of a 2x2 transpose convolution, followed by a concatenation, two 3x3 convolutions, and a ReLU activation function. 

\subsection{Similarity Fusion Blocks}
\label{subsectionsection:crossblockmodule}
The Similarity Fusion is a technique aimed at capturing cross-dimensional dependencies in feature maps and handling datasets with inconsistent labels. 
This approach effectively models complex relationships across input dimensions, facilitating improved representation learning and feature extraction by exploiting correlations between spatial, temporal, or channel-wise relationships.
Unlike the standard fusion module \cite{huang2024dra}, which achieves feature fusion through pixel-wise summation or channel-wise concatenation, the similarity fusion block uses supervision and selection similarity knowledge to reduce irrelevant and noisy signals in the output. This is crucial for capturing the synergistic potential of diverse datasets from multiple modalities, encompassing different organs with various diseases, and for mitigating negative knowledge transfer during training.
Given an input image, we enhance its quality and remove noise by applying a Gaussian filter \cite{ito2000gaussian} at various smoothing rates using different sigma values, producing three new images. To further reduce the noise, we use the Euclidean distance similarity measure \cite{wang2005euclidean} at the pixel level to calculate the similarity. 
Pixels from the same position on all three images are grouped together. Each group contains three pixels, one from each of the three different feature maps. The pixel similarity is measured at the group level. Within each group, the pixel that is most similar to the other two is chosen, while the other two are excluded. The similarity is measured by finding the pixel with the shortest distance to the other two. 
 The similarity fusion block is integrated into the network's architecture before and after every convolutional block in both the contracting and expanding paths. It is also used in the bridge layer. 
 This innovation captures image-specific information while ensuring that only common or similar information across all image samples is used to rebuild the feature map. As the feature map move through the convolutional blocks, dissimilar information is progressively discarded, thereby removing irrelevant knowledge and mitigating the problem of negative knowledge transfer.
A high level diagram to demonstrate the similarity block is shown at the bottom of Figure ~\ref{fig:MMIS-Net} \hhl{ and a snippet of the similarity fusion pseudocode is shown in Algorithm~ \ref{fig:similarity_fusion}.}

\begin{algorithm*}
\caption{Snippet of the Similarity Fusion Pseudocode}
\begin{algorithmic}[1]
\For{each fusion map}
    \State Generate three fusion maps at different smoothing scales
    \For{each pixel}
        \For{each position along the Z-axis}
            \State Compute the similarity between pixels using the distance matrix
            \State Select the two pixels with the shortest distance
            \State Fuse the selected pixels across the Z-axis
        \EndFor
    \EndFor
\EndFor
\end{algorithmic}
\label{fig:similarity_fusion}
\end{algorithm*}

\subsection{The Residual Connection}
\label{subsectionsection:residualconnection}
Residual connection \cite{he2016deep} is a skip connection that enables the network to learn residual mappings instead of directly fitting the desired underlying mapping. Traditional deep networks aim to approximate the underlying mapping \( H(x) \) using stacked layers. However, during training, it can be challenging for deeper networks to learn these mappings effectively. Residual learning introduces the concept of learning residual functions, denoted as \( F(x) = H(x) - x \), where \( H(x) \) is the desired mapping and \( x \) is the input to a certain layer.
The residual connection is incorporated into the network's architecture at every level in both the contracting and expanding paths to mitigate the problem of vanishing gradients.

\subsection{The Class-adaptive Loss Function}
\label{subsectionsection:classadaptiveloss}
The loss function used is a combination of cross-entropy and Dice loss. We employed binary cross-entropy loss and a modified Dice loss. The regular dice loss is calculated individually for each image in a batch, whereas we jointly calculate the dice loss for all images in the input batch. This approach helps regularize the loss when only a few voxels of one class appear in one image, while a larger area is present in another image of the same batch. Consequently, inaccurate predictions of a few pixels in one image have a limited impact on the overall loss.

Between the contracting and expanding paths is a bridge layer composed of a  similarity fusion block  to ensure a smooth transition from one path to the other. At the end of the expanding path is a classification layer to classify each pixel as belonging to the background or one of the segmented classes.

\section{Experiments}
\label{section:experiments}

\subsection{Dataset}
\label{subsectionsection:dataset}

\begin{figure*}[h]
\centerline{\includegraphics[width=\textwidth]{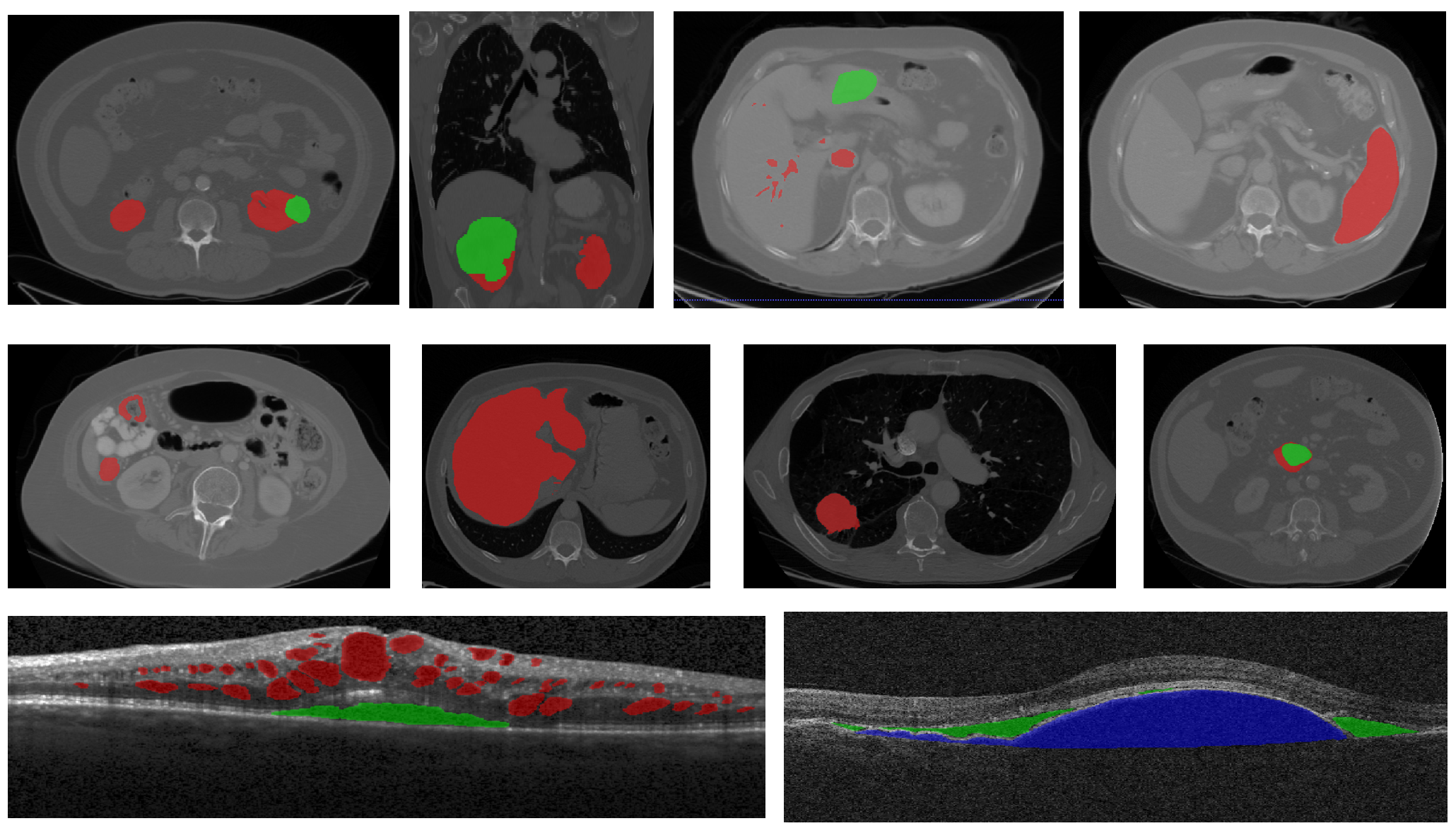}}
\caption{An illustration of B-Scans from different datasets, showcasing various organs, modalities, and diseases, highlighting the high diversity of the datasets. The first row presents examples of CT B-scans of an affected kidney from various views, with annotated segments highlighting the kidney and tumor. The second row features CT B-scans of an affected lung from different perspectives, with annotations marking the lung and tumor regions. Finally, the third row displays OCT B-scans of an affected retina, showing annotated layers and three distinct fluid regions.}
\label{fig:dataset}
\end{figure*}
A total of 10 datasets originating from the Medical Segmentation Decathlon (MSD) \cite{medicaldecathlon}, Pelvis \cite{synapse}, Pancreas CT \cite{nihPancreasCT}, KiTS19 \cite{kits19}, and RETOUCH \cite{bogunovic2019retouch}, datasets were used in this work. The datasets were annotated for 19 anatomic structures, consisting of 1337 volumes across 2 modalities: computed tomography (CT) and optical coherence tomography (OCT). These datasets cover 19 segmentation tasks and one detection task. The MSD datasets used are as follows:
Liver: This dataset consists of 201 contrast-enhanced CT images from patients with primary cancers and metastatic liver disease. The segmented regions of interest are the liver and tumors inside. It was acquired at the IRCAD Hopitaux Universitaires, Strasbourg, France.
Pancreas: This dataset consists of 421 CT scans of patients undergoing resection of pancreatic masses. The segmented regions of interest are the pancreatic parenchyma and pancreatic mass (cyst or tumor). It was acquired at the Memorial Sloan Kettering Cancer Center, New York, US.
Hepatic Vessels: This dataset consists of 443 CT scans of patients with a variety of primary and metastatic liver tumors. The segmented regions of interest are the vessels and tumors within the liver. It was acquired at the Memorial Sloan Kettering Cancer Center, New York, US.
Lung: This dataset consists of 96 CT scans of patients with non-small cell lung cancer, and the segmented region of interest is the lung tumors. It was collected from the Cancer Imaging Archive \cite{cancerimagingarchive}.
Spleen: This dataset consists of 61 CT scans of patients undergoing chemotherapy treatment for liver metastases, and the segmented region of interest is the spleen. It was acquired at the Memorial Sloan Kettering Cancer Center, New York, USA.
Colon: This dataset consists of 190 CT scans of patients undergoing resection of primary colon cancer, and the segmented region of interest is the primary colon cancer. It was acquired at the Memorial Sloan Kettering Cancer Center, New York, USA.
KiTS19 \cite{kits19}: This dataset consists of 300 CT scans. The segmented regions of interest are the kidneys and kidney tumors. They were acquired at the University of Minnesota Medical Center, USA.
Pelvis \cite{synapse}: This dataset consists of 50 CT scans, and the segmented regions of interest are the uterus, bladder, rectum, and bowel. The dataset was acquired from the Vanderbilt University Medical Center (VUMC), USA, and the Erasmus Medical Center (EMC) Cancer Institute in Rotterdam, the Netherlands.
Pancreas CT \cite{nihPancreasCT}: This dataset consists of 82 CT scans, and the segmented region of interest is the pancreas. The dataset was acquired from the National Institutes of Health \cite{nihPancreasCT}.
RETOUCH \cite{bogunovic2019retouch}: This dataset consists of 112 retinal optical coherence tomography (OCT) scans of patients with early age-related macular degeneration (AMD) and diabetic macular edema (DME), collected from three device vendors: Cirrus, Spectralis, and Topcon. For a fair comparison, the training set consisting of 70 scans is available to the public, and the testing set consisting of 42 hidden scans is held by the organizers. Submission and evaluation of predictions on the testing dataset are arranged privately with the organizers, and the results are sent to the participants. The dataset was segmented for three regions of interest: intraretinal fluid (IRF), subretinal fluid (SRF), and pigment epithelium detachments (PED). The dataset was acquired from the Medical University of Vienna (MUV) in Austria, Erasmus University Medical Centre (ERASMUS), and Radboud University Medical Centre (RUNMC) in the Netherlands.
Examples of the datasets are shown in Figure ~\ref{fig:dataset}, and further details about the datasets' composition are provided in Table ~\ref{tab:datasetcomposition}.

\begin{table*}[t!]
\small
\addtolength{\tabcolsep}{-4pt}
\centering
\begin{tabular}{l c c c c c c}
\toprule\toprule

Datasets & Modality & Labels  &  Training   & Shape  & Spacing [mm] \\
\midrule
 
Liver \cite{medicaldecathlon}   & CT & Liver, L. Tumor    & 131    & 432x512x512 &  (1, 0.77, 0.77)  \\ 

Lung \cite{medicaldecathlon}   & CT & Lung nodules    & 63    & 252x512x512 &  (1.24, 0.79, 0.79)  \\ 

Pancreas \cite{medicaldecathlon}   & CT & Pancreas, P. Tumor   & 281    & 93x512x512 &  (2.5, 0.80, 0.80)  \\ 

H. Vessels \cite{medicaldecathlon}   & CT & H. vessels, H. Tumor   & 303    & 49x512x512 &  (5, 0.80, 0.80)  \\ 

Spleen \cite{medicaldecathlon}   & CT &  Spleen   & 41    & 90x512x512 &  (5, 0.79, 0.79)  \\ 

Colon \cite{medicaldecathlon}   & CT &  Colon cancer  & 126    & 95x512x512 &  (5, 0.78, 0.78)  \\ 

Pelvis \cite{synapse}   & CT &  Ut, Bl, Rec, Bow  & 30    & 180x512x512 &  (2.5, 0.98, 0.98)  \\ 

Pancreas CT \cite{nihPancreasCT}   & CT &  Pancreas  & 82    & 217x512x512 &  (1, 0.86, 0.86)  \\ 

KiTS19  \cite{kits19}   & CT &  Kidney, K.Tumor  & 210    & 107x512x512 &  (3, 0.78, 0.78)  \\ 

RETOUCH  \cite{bogunovic2019retouch}   & OCT & IRF, SRF, PED  & 70    & 128 x512x512 &  (0.01, 0.01, 0.05)  \\ 

\textbf{Total}  &   &    & \textbf{1337 }     &       &    \\ 
 \\ 
\end{tabular}
\caption{Summary table of the datasets used, showing the modalities, anatomic structures, number of training cases, median shapes, and image spacings. The abbreviations used in this table are L. Tumor, Liver Tumor;
P. Tumor , Pancreas Tumor;
H. Vessels, Hepatic Vessels;
H. Tumor,  Hepatic Tumor;
Ut, Uterus; Bl, Bladder; Rec, Rectum; and Bow, Bowel.}
\label{tab:datasetcomposition}
\end{table*}

\subsection{Training and Testing}
All datasets were combined into a one-hot label space as demonstrated in Table~\ref{tab:labels}. This approach effectively handles annotations present in one dataset but missing in another. For instance, in this work, there are two different pancreas datasets: \cite{medicaldecathlon}, which includes segmentations for the pancreas and pancreas tumor, and \cite{nihPancreasCT}, which includes segmentations only for the pancreas. The one-hot label space efficiently separates these as different labels without overlap. During training, MMIS-Net leverages the synergistic potential of one dataset to improve the performance of the other and vice versa. It also supports overlapping target structures, such as vessels or cancer classes within an organ, and retains different annotation protocol characteristics for the same target structure.
During training, the following parameters were used: the learning rate was set to 0.1, the optimizer was Adam \cite{diederik2014adam}, the maximum epoch was set to 1000, the sigma parameters were fixed, and early stopping was used to avoid overfitting. The loss function used was a combination of cross-entropy and Dice loss. 
In this work we aimed to improve the segmentation and detection performance on retinal OCT fluids. For this, we trained the algorithm by combining the 1337 publicly available volumes of the training sets of all 10 datasets and evaluated the results on the hidden test set of the RETOUCH
\cite{bogunovic2019retouch} dataset set. Three evaluation metrics were used:
Dice Score (DS): This measures the overlap between the predicted and ground truth segments, calculated as twice the intersection divided by the union. It ranges from 0 to 1, with 1 being the perfect score and 0 being the worst. In clinical settings, DS is essential for assessing how well a model can capture the exact shape and boundary of structures such as diseases, or lesions. High DS values suggest the segmentation closely aligns with expert annotations, making it reliable for clinical use.
Absolute Volume Difference (AVD): This is the absolute difference between the predicted and ground truth volumes. The value ranges from 0 to 1, with 0 being the best result and 1 being the worst. In clinical settings, accurate disease/fluid volume measurement is critical in treatments like radiotherapy, where the dose is calculated based on fluid volume. A low AVD means the model can accurately estimate volume, ensuring that treatment plans and dosages are based on precise measurements.
Area Under the Curve (AUC): This measures the ability of a binary classifier to distinguish between classes. The AUC score ranges from 0 to 1, with 1 being the perfect score and 0 being the worst. In clinical settings, for early disease detection, a high AUC is crucial as it reflects the model's ability to distinguish even subtle differences between healthy and abnormal tissue. This distinction is valuable in preventive care and early intervention, where the cost of a missed detection is high. Also, a high AUC score suggests that the model is consistently able to distinguish between target and background across diverse data, increasing its reliability for clinical application in real world scenarios.

The DS and AVD were used to evaluate the segmentation of the retinal fluids on OCT scans, while the AUC was used to evaluate the detection of fluids on the retinal OCT scans. 
For fair comparison, we used the DS, AVD, and AUC evaluation metrics as they were the same evaluation metrics used by the organizers of the RETOUCH grand challenge for the retinal OCT dataset. While Dice Score (DS) and Absolute Volume Difference (AVD) measure overlap and volume estimation performance, AUC offers a broader assessment of pixel detection and classification between classes. In clinical evaluations, combining AUC with DS and AVD provides a well-rounded view of a model’s performance, ensuring that it not only accurately segments but also effectively differentiates between relevant and non-relevant areas. Submission is privately organized and sent to the organizers, and the results are emailed to the teams. Submissions are limited to a maximum of three per team hence. The experimental setup was the same for all the experiments. The algorithm was written in Python using PyTorch backend libraries.

\begin{table}[h!]
\centering
\begin{tabular}{l l}
\toprule\toprule
Assigned Value & Region  \\
  \midrule

0     & Background        \\
1     & Liver        \\
2     & Liver tumor        \\
3     & Pancreas       \\
4     & Pancreas tumor      \\
5     & Hepatic vessels       \\
6     & Hepatic vessels  tumor     \\
7     & Lung tumor       \\
8     & Spleen      \\
9     & Colon cancer      \\
10    & Bladder      \\
11    & Ulterus      \\
12    & Rectum      \\
13    & small bowel      \\
14    & Pancreas      \\
15    & Kidney      \\
16    & Kidney tumor      \\
17    & Intraretinal Fluid (IRF) \\
18    & Subretinal Fluid (SRF) \\
19    & Pigment Epithelium Detachments (PED) \\

\\
\end{tabular}
\caption{Evaluation performance of the fluids detection, measured in Area Under the Curve (AUC), grouped by segmented classes with their averages in columns and teams in rows on the hidden test set of the RETOUCH grand challenge.}
\label{tab:labels}
\end{table}

The models were trained on a GPU work station with NVIDIA RTX A5000 48GB and took 14 hours to train. The models were implemented in Python, using PyTorch library.

\subsection{Results}
The model was validated on a hidden (or blind) Retouch test dataset, simulating a real-world scenario, with data acquired from three different sources or vendor machines (Topcon, Spectralis and Cirrus). Based on the experimental results, we observed the following:
\begin{enumerate}
\item The MMIS-Net outperformed the SOTA algorithms on the segmentation task with a clear improvement in both DS and AVD, obtaining a mean of 0.83 and 0.035, respectively, on the RETOUCH retinal OCT hidden test set.

\item The MMIS-Net obtained the best DS score in all three fluid classes and the best AVD in two out of the three classes for the segmentation task on the RETOUCH retinal OCT hidden test set.

\item  The MMIS-Net achieved a perfect AUC score of 1 alongside two other SOTA algorithms for the detection task on the RETOUCH retinal OCT hidden test set.

\item MMIS-Net, outperformed SAMedOCT, a large foundation model for medical image segmentation while using fewer resources. SAMedOCT was trained for 20 hours on an NVIDIA A100, 80GB GPU workstation, while MMIS-Net was trained for 14 hours on an NVIDIA RTX A5000, 48GB GPU workstation.

\item  SAMedOCT obtained the best AVD of 0.033 for the segmentation of the PED fluid on the RETOUCH retinal OCT hidden test set.

\item For the RETOUCH retinal OCT segmentation and detection tasks, as well as the segmentation task, we notice a constant and steady high performance of the MMIS-Net algorithm, highlighting its robustness and generalizability.

\end{enumerate}

High DS values indicate that the segmentation closely aligns with human expert annotations, while low AVD shows that the model can accurately estimate volume, allowing treatment plans and dosages to be based on precise measurements. MMIS-Net outperformed other state-of-the-art (SOTA) algorithms by a clear margin in DS score across all classes and in AVD across all classes except for the PED class. These results demonstrate MMIS-Net's ability to accurately capture the shape and structure of retinal fluids/diseases (as reflected in DS) and to precisely measure their volume (as reflected in AVD) in a clinical context

\begin{figure*}[h!]
\centerline{\includegraphics[width=\textwidth]
{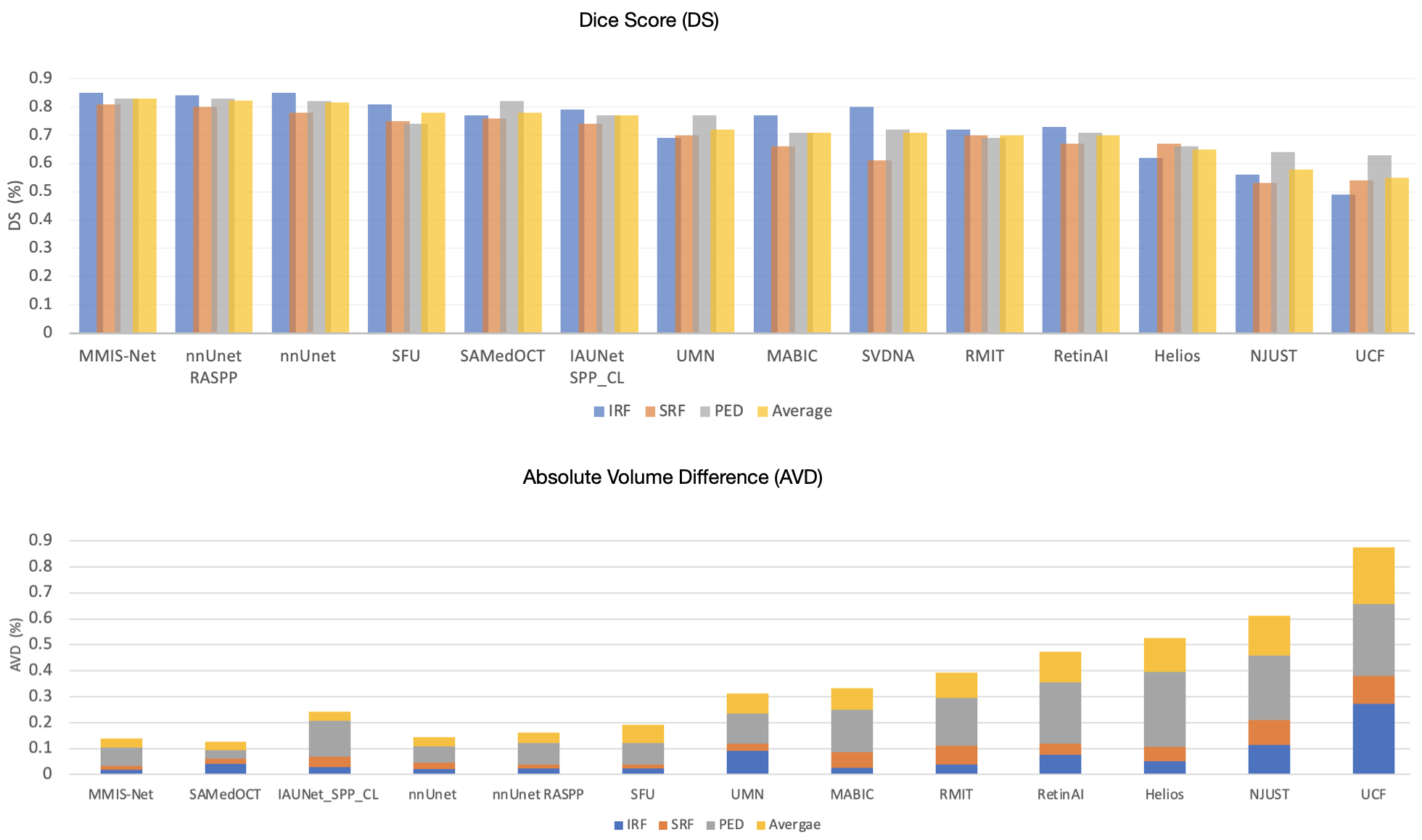}}
\caption{Comparison of performance evaluations for methods/teams, categorized by segmented classes and averages (Avg.), on the hidden test set of the RETOUCH grand challenge, measured with Dice Score (DS) and Absolute Volume Difference (AVD), presented in bar charts. High DS values indicate that the segmentation closely aligns with human expert annotations, while low AVD shows that the model can accurately estimate volume, allowing treatment plans and dosages to be based on precise measurements. MMIS-Net outperformed state-of-the-art (SOTA) algorithms in every class for both DS and AVD metrics, except for the PED class in AVD, where SAMedOCT achieved the best score of 0.035. Additionally, MMIS-Net and nnU-Net jointly achieved the highest DS score for the IRF class.}
\label{fig:ds_avd}
\end{figure*}

\begin{figure*}[h!]
\centerline{\includegraphics[width=\textwidth]
{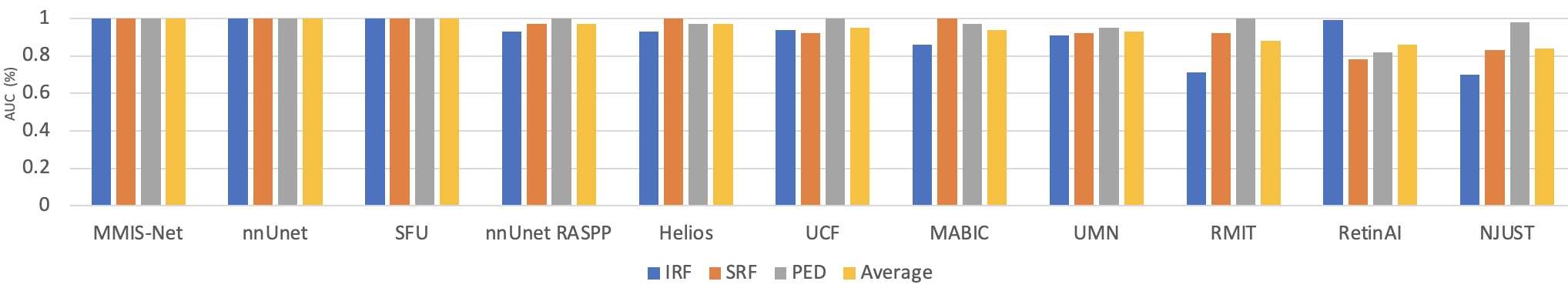}}
\caption{Performance evaluation of fluid detection, measured by Area Under the Curve (AUC), categorized by segmented classes and their averages, and grouped by teams on the hidden test set of the RETOUCH grand challenge. A high AUC indicates the model's ability to distinguish even subtle differences between healthy and abnormal tissue, which is valuable for early detection and intervention in retinal diseases. MMIS-Net achieved a perfect AUC of 1 in every single class, alongside two other state-of-the-art (SOTA) algorithms: nnU-Net and SFU}.
\label{fig:auc}
\end{figure*}

\begin{figure*}[h]
\centerline{\includegraphics[width=\textwidth]
{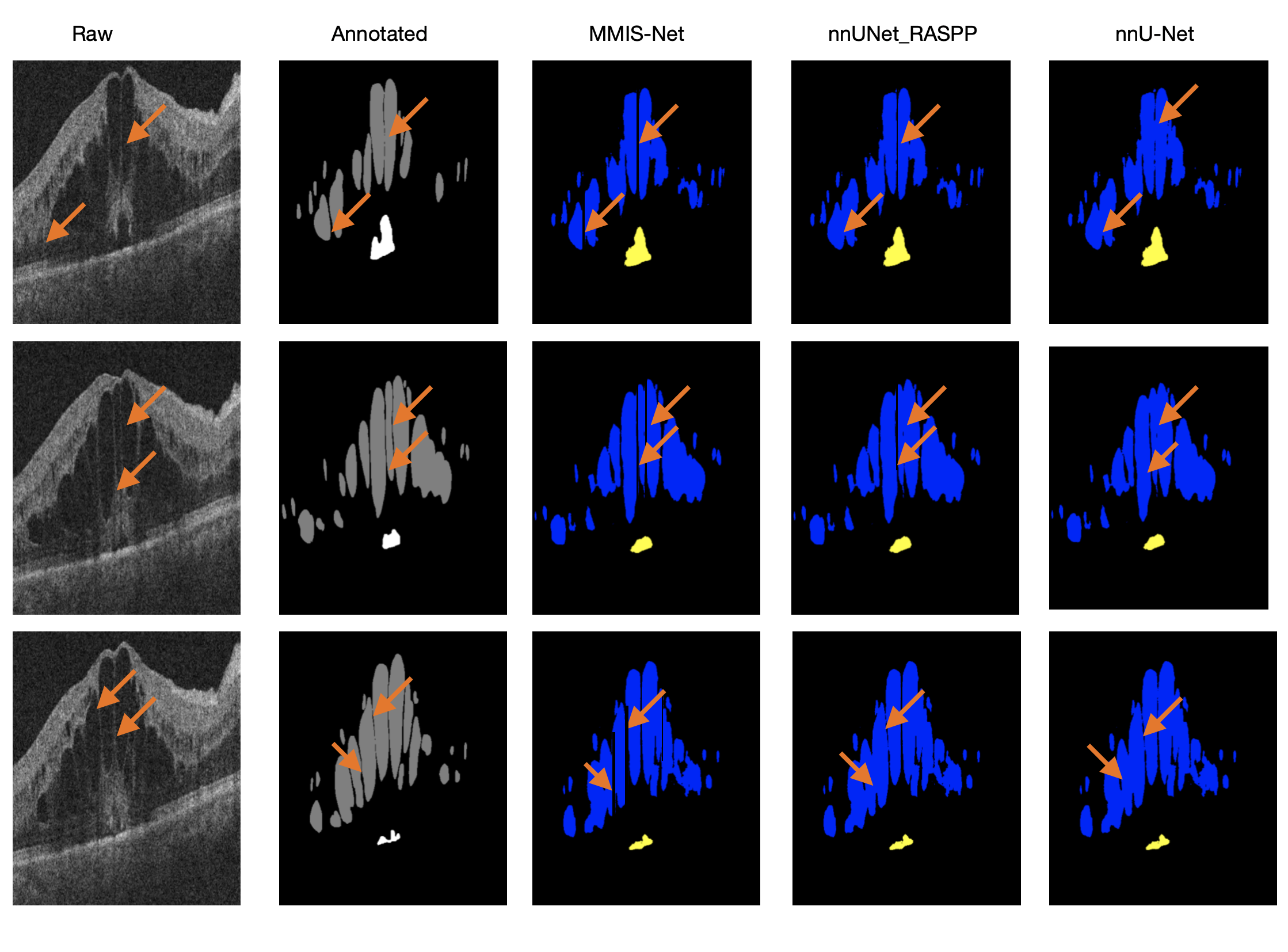}}
\caption{A visualization comparison of predicted output demonstrating the superior performance of MMIS-Net over SOTA algorithms (nnUNet\_RASPP and nnU-Net)  on the training set of the Retouch dataset using a 5-fold cross-validation. Orange arrows highlight the details captured or missed by the models. From left to right, the images display the raw original image, the ground truth (annotations from a human expert), and the predicted outputs from MMIS-Net, nnUNet\_RASPP, and nnU-Net. In the first row, vertical lines are observed cutting across both the raw images and the ground truth. These lines were clearly detected by MMIS-Net and nnUNet\_RASPP, but nnU-Net detected only one of the lines. In the second row, for a different scan, two vertical lines visible in the raw image were annotated in the ground truth. While MMIS-Net segmented both lines, nnUNet\_RASPP and nnU-Net managed to segment only one line. Finally, in the third row, similar patterns were observed: two vertical lines, clearly visible in the raw and annotated images, were segmented by MMIS-Net, but nnUNet\_RASPP and nnU-Net again segmented only one of the lines. This demonstrates MMIS-Net's superior ability to accurately detect and segment key features compared to the other models}.
\label{fig:visualisation}
\end{figure*}

Segmentation measured in DS and AVD on the RETOUCH retinal OCT hidden test set is highlighted in Table~\ref{tab:dsavd}, and the detection task measured in AUC is highlighted in Table~\ref{tab:results_auc}, with their corresponding bar charts in Figure~\ref{fig:ds_avd} and Figure~\ref{fig:auc}, respectively. To further demonstrate the high performance of the MMIS-Net, a visualization comparison of the predicted output of 5-fold cross validation on the RETOUCH training dataset is demonstrated in Figure~\ref{fig:visualisation}.

\begin{table*}[h!]
\small
\addtolength{\tabcolsep}{0pt}
\centering
\begin{tabular}{l|cccc|cccc }
\hline
\hline
\multirow{3}{*}{Methods/Teams}   & \multicolumn{4}{c|}{Dice Score (DS)} 

& \multicolumn{4}{c}{Absolute Volume Difference (AVD)}  \\

 & IRF & SRF & PED & Avg. & IRF & SRF & PED & Avg.   \\
\hline

\textbf{MMIS-Net} 
& \textbf{0.85} & \textbf{0.81} & \textbf{0.83} & \textbf{0.83}
& \textbf{0.018} & \textbf{0.015}  & 0.071 & \textbf{0.035}   \\

nnUNet\_RASPP & 
0.84 & 0.80 & \textbf{0.83} & 0.82 
& 0.023 & 0.016  & 0.083  & 0.041   \\

nnU-Net & 
\textbf{0.85} & 0.78 & 0.82  & 0.81
& 0.019 & 0.017  & 0.074  & 0.036   \\

SFU & 
0.81 & 0.75 & 0.74  & 0.78
& 0.030 & 0.038  & 0.139  & 0.069   \\

SAMedOCT & 
0.77 & 0.76 & 0.82  & 0.78
& 0.042 & 0.020  & \textbf{0.033}  & 0.032   \\

IAUNet\_SPP\_CL & 
0.79 & 0.74 & 0.77  & 0.77
& 0.021 & 0.026  & 0.061  & 0.036   \\

UMN & 
0.69 & 0.70 & 0.77  & 0.72
& 0.091 & 0.029  & 0.114  & 0.078   \\

MABIC & 
0.77 & 0.66 & 0.71  & 0.71
& 0.027 & 0.059  & 0.163  & 0.083   \\

SVDNA & 
0.80 & 0.61 & 0.72  & 0.71
& -- & --  & --  & --   \\

RMIT & 
0.72 & 0.70 & 0.69  & 0.70
& 0.040 & 0.072 & 0.182  & 0.098   \\

RetinAI & 
0.73 & 0.67 & 0.71  & 0.70
& 0.077 & 0.041 & 0.237  & 0.118   \\

Helios & 
0.62 & 0.67 & 0.66  & 0.65
& 0.051 & 0.055 & 0.288  & 0.132   \\

NJUST & 
0.56 & 0.53 & 0.64  & 0.58
& 0.113 & 0.096 & 0.248  & 0.153   \\

UCF & 
0.49 & 0.54 & 0.63  & 0.55
& 0.272 & 0.107 & 0.276  & 0.219   \\

\newline
\end{tabular}

\caption{Performance evaluations of methods/teams, grouped by segmented classes and averages (Avg.), on the hidden test set of the RETOUCH grand challenge, measured in Dice Score (DS) and Absolute Volume Difference (AVD).}
\label{tab:dsavd}
\end{table*} 

\begin{table*}[h!]
\addtolength{\tabcolsep}{18pt}
\centering
\begin{tabular}{l c c c c c}
\toprule\toprule
Methods & IRF  &  SRF  & PED  & Avg.  \\
  \midrule

\textbf{MMIS-Net}      &\textbf{1.00}  & \textbf{1.00}    & \textbf{1.00}    & \textbf{1.00}     \\ 

nnU-Net      &\textbf{1.00}  & \textbf{1.00}    & \textbf{1.00}    & \textbf{1.00}     \\ 
SFU         &\textbf{1.00}  & \textbf{1.00}    & \textbf{1.00}    & \textbf{1.00}     \\

nnUNet\_RASPP     & 0.93      & 0.97   & \textbf{1.00}    & 0.97    \\

Helios            & 0.93     & \textbf{1.0}    & 0.97     & 0.97     \\ 

UCF               & 0.94    & 0.92        & \textbf{1.00}    & 0.95    \\  

MABIC              & 0.86     & \textbf{1.00}      & 0.97    & 0.94      \\

UMN                & 0.91     & 0.92       & 0.95     & 0.93      \\

RMIT                & 0.71     & 0.92      & \textbf{1.0}     & 0.88     \\

RetinAI             & 0.99     & 0.78       & 0.82     & 0.86      \\   

NJUST             & 0.70     & 0.83       & 0.98    & 0.84        \\ 
\\
\end{tabular}
\caption{Evaluation performance of the fluids detection, measured in Area Under the Curve (AUC), grouped by segmented classes with their averages in columns and teams in rows on the hidden test set of the RETOUCH grand challenge.}
\label{tab:results_auc}
\end{table*}

\newpage
\hhl{Figure \ref{fig:visualisation} presents a visual comparison of MMIS-Net against state-of-the-art (SOTA) algorithms (nnUNet\_RASPP and nnU-Net), demonstrating its superior performance. In all three rows, as indicated by the orange arrows, clear visible lines in the raw and annotated datasets were accurately detected by MMIS-Net, whereas nnUNet\_RASPP and nnU-Net struggled to capture these lines.}

\section{Conclusions}
\label{section:conclusions}
In this work, we propose MMIS-Net, a novel algorithm designed to segment multiple lesions from various organs across diverse image modalities using a single model. To address the issue of negative knowledge transfer, MMIS-Net introduces Similarity Fusion Blocks within its architecture. These blocks utilize supervision and selection knowledge transfer for feature map fusion at the pixels level, effectively reducing irrelevant and noisy signals in the output. Additionally, we efficiently created a one-hot label space to address the inconsistent class definitions and label contradiction problem from diverse modalities and body regions. 

The MMIS-Net was evaluated on the hidden test set of the RETOUCH grand challenge, outperforming and state-of-the-art (SOTA) and SAMedOCT, a large foundation models for medical image segmentation  algorithms while using fewer resources. SAMedOCT was trained for 20 hours on an NVIDIA A100, 80GB GPU  workstation, while MMIS-Net was trained for 14 hours on an NVIDIA RTX A5000 GPU workstation.
MMIS-Net achieved a mean Dice score (DS) of 0.83 and an absolute volume difference (AVD) of 0.035 for the retinal fluids segmentation task, and a perfect Area Under the Curve (AUC) of 1 for the fluid detection task.

We believe that the model's superior fluid segmentation and detection performance, is due to the integration of the following key features into the CNN backbone: 1) Similarity Fusion blocks for supervision and similarity knowledge selection for feature map fusion, 2) a one-hot label space to address inconsistent class definitions and label contradictions, handling classes absent in one dataset but annotated in another, while retaining different annotation protocol characteristics for the same target structure during training, and 3) residual connections to combat the problem of vanishing gradients.

The performance and generalizability of MMIS-Net suggest that it can contribute to improved clinical outcomes and diagnostic capabilities by: (i) aiding in the early detection or diagnosis of cases by providing clinicians with a valuable second opinion, serving as a reliable decision-support tool, (ii) handling less complex tasks, allowing clinicians to focus on more complex cases, thereby saving time, and (iii) enabling early diagnosis, which can save lives, reduce costs, and alleviate the socio-economic burden on both patients and the healthcare system. Furthermore, once trained, the model is lightweight and can be deployed without requiring significant computational resources or specialized expertise.

\subsection{Limitation}
The limitations of this approach are as follows:\\
\hhl{1) The algorithm was validated using hidden cases from the Retouch grand challenge dataset, which participants do not have access to. According to the challenge rules, each participant is allowed a maximum of 3 submissions to ensure a fair comparison, which restricts opportunities for conducting additional statistical tests.}\\
2) The inter-observer agreement score among annotators allows us to compare our results with the level of agreement among human experts. However, in the Retouch dataset, inter-observer agreement data is not available. \\
\hhl{3) Certain hyperparameters were set manually, which may not yield the optimal model performance. One possible solution is to use an enhanced self-parameterized pre-processing approach of the nnU-Net \cite{ronneberger2015u}, provided sufficient computational resources are available.}

\subsection{Future Research }
In this work, MMIS-Net has so far been evaluated only on the RETOUCH challenge dataset. To further demonstrate its generalization performance, in the future we plan to participate in more medical image segmentation challenges and evaluate the model on diverse datasets spanning various diseases, organs, and imaging modalities in the future. \hhl{More specifically, we aim to optimize the model's time and space complexity to enable scalability, allowing for an increased dataset size from other sources without requiring additional computational resources. This, will in turn, enhances the diversity of the training datasets.}

\subsection*{Disclosures}
The authors have no relevant financial interests in the paper and no other potential conflicts of
interest to disclose.

\subsection* {Code and  Data Availability} 
The datasets used for this study are publicly available (see details in Section 4.1). The code used for this study is not publicly accessible but may be provided to qualified researchers
upon reasonable request to the corresponding author.

\subsection* {Acknowledgments}
We would like to express our sincere gratitude to Hrvoje Bogunovic for his invaluable support and advice during our participation in the RETOUCH competition.


\bibliography{bibliography}   
\bibliographystyle{spiejour}   




\listoffigures
\listoftables

\end{spacing}
\end{document}